\documentclass[prb,preprint]{revtex4-1}
\usepackage[utf8]{inputenc}
\usepackage[utf8]{inputenc}
\usepackage{siunitx}
\usepackage{amsmath,amssymb,amsfonts,mathtools,xcolor}
\usepackage{graphicx}

\newcommand{\ra}{\rightarrow}

\newcommand{\tx}{{\tilde x}}
\newcommand{\tr}{{\tilde r}}
\newcommand{\bomega}{{\boldsymbol{\omega}}}

\begin{document}

\title{Non-relativistic Geometry and the Equivalence Principle}
\author{Anton Kapustin}
\email{kapustin@theory.caltech.edu}
\affiliation{California Institute of Technology, Pasadena, CA 91125}
\author{Marc Touraev}
\email{touraevm@iu.edu}
\affiliation{Physics Department, Indiana University, Bloomington, IN 47405}
\date{August 2020}

\begin{abstract}
    We describe a geometric and symmetry-based formulation of the equivalence principle in non-relativistic physics. It applies both on the classical and quantum levels and states that the Newtonian potential can be eliminated in favor of a curved and time-dependent spatial metric. It is this requirement that  forces the gravitational mass to be equal to the inertial mass. We identify the symmetry responsible for the equivalence principle as the remnant of time-reparameterization symmetry of the relativistic theory. We also clarify the transformation properties of the Schr\"odinger wave-function under arbitrary frame changes.
\end{abstract}

\maketitle

\section{Introduction}

The Strong Equivalence Principle states that all physical effects of a uniform static gravitational field can be eliminated by going to a uniformly accelerated frame. This implies the universality of free fall (the gravitational acceleration of all bodies in a uniform static gravitational field is the same), which is one of the best tested physical laws (see e.g. \cite{torsionbalanceWEP,atomicWEP}). The Strong Equivalence principle also motivated A. Einstein to create General Relativity Theory. Within General Relativity Theory, gravity is reinterpreted as a curved geometry of space-time, and a uniform static gravitational field corresponds to a flat space-time. The Strong Equivalence Principle then becomes obvious.

On the other hand, the theoretical status of the Strong Equivalence Principle and the universality of free fall in non-relativistic mechanics is not widely known. This is because gravity is described by a non-geometric quantity, the Newtonian potential. There seems to be no reason why the gravitational mass (that is, the strength of the coupling to the Newtonian potential) must be the same as the inertial mass (the coefficient in the kinetic energy). In non-relativistic quantum mechanics even the formulation of the equivalence principle is not completely settled, see \cite{quantumEP} and references therein.

A popular geometric framework for non-relativistic physics is the Newton-Cartan geometry introduced by E. Cartan \cite{Cartan1,Cartan2}. The Newton-Cartan formalism is covariant with respect to arbitrary diffeomorphisms of space-time as well as ``Milne boosts", see e.g. \cite{Geracie} for a recent review. However, neither of these  symmetries enforces the equivalence principle. Rather, the equivalence principle originates from a certain non-geometric local symmetry which is not required by the Newton-Cartan geometry. This symmetry was noticed already by K. Kucha\v{r} \cite{Kuchar} and is well-known to experts on the Newton-Cartan formalism (see e.g.  \cite{DuvalKunzle,Bekaert&Morand,Obersetal}). The origin of this symmetry and its relationship with the equivalence principle is also part of the folklore, but as far as we know it has not been explicitly stated in the published literature. 

In this short note,  we present a simplified treatment of the equivalence principle in non-relativistic physics.  We show how the non-geometric symmetry enforcing the equivalence principle arises from the  time-reparameterization symmetry of the relativistic theory in the limit $c\ra\infty$.  Then we formulate the non-relativistic equivalence principle in a physically transparent form which takes into account both the Newtonian potential and gravitational waves. We use a simplified version of the Newton-Cartan formalism which does not require considering Milne boosts. While none of these results are completely new, we hope that our exposition makes the underlying physics accessible to a wider physics audience. We also believe that the simplified Newton-Cartan approach deserves to be better known, since it might be useful in other problems, such as the interaction of gravitational waves with quantum matter and Effective Field Theories of hydrodynamics.

\section{Non-relativistic geometry as a foliated geometry}

In non-relativistic physics, clocks can be synchronized instantaneously and thus the notion of simultaneity has an invariant meaning. Thus the space-time is foliated by codimension-1 submanifolds of simultaneous events (that is, spatial slices). The natural invariance group of non-relativistic physics is the group of foliated diffeomorphisms, that is, diffeomorphisms which preserve every spatial slice (that is, every leaf of the foliation). To describe this group more explicitly, we choose a time coordinate $t$ (that is, a parameterization of the space of leaves of the foliation) as well as local coordinates $x^k, k=1,2,3,$ on every spatial slice. Then there are two kinds of allowed coordinate transformations: time reparameterizations $t\mapsto \tilde t=f(t)$, and time-dependent changes of spatial coordinates:
\begin{equation}
\label{framechange}
x^k\mapsto \tx^k=\tx^{k}({\bf x}, t),\quad k=1,2,3.
\end{equation}
Since $\tx^k$ is allowed to depend on $t$, in general the latter transformation is not just a coordinate change, but a reference frame change. In what follows we will assume that a global time coordinate has been chosen and focus on ensuring covariance with respect to transformations (\ref{framechange}). 

Under transformations (\ref{framechange}) partial derivatives with respect to time and space coordinates transform  as follows:
\begin{equation}
\frac{\partial}{\partial \tx^k}=\left(A^{-1}\right)^j_k\frac{\partial}{\partial x^j},\quad \left. \frac{\partial}{\partial t}\right|_\tx=\left.\frac{\partial}{\partial t}\right|_x-(A^{-1})^{k}_{j}B^j\frac{\partial}{\partial x^k},
\end{equation}
where
\begin{equation}
A^{j}_k=\frac{\partial \tx^{j}}{\partial x^{k}},\quad 
B^j= \frac{\partial \tx^{j}}{\partial t}.
\end{equation}
Since the time-derivative transforms non-covariantly, to write covariant actions and equations of motion one needs to introduce a suitable connection. As explained in more detail below, a connection is locally encoded by three functions $N^j({\bf x},t), j=1,2,3$ which under a frame change (\ref{framechange}) transform as follows:
\begin{equation}
\label{transformation}
{\tilde N}^{j}=A^{j}_k N^k-B^{j}.
\end{equation}

In the mathematical literature such an object defines what is known as an Ehresmann connection \cite{KN}. In general, an Ehresmann connection gives a notion of parallel transport for a fiber bundle $\pi : E \rightarrow S$ with a typical fiber $F$. Namely, given a path on $S$ connecting $s_0\in S$ and $s_1\in S$, an Ehresmann connection provides a diffeomorphism (smooth 1-1 idenfitication) of the fibers over $s_0$ and $s_1$. This is the only reasonable notion of parallel transport when the fiber does not have any additional structure beyond that of a smooth manifold. Connections on principal $G$-bundles and affine connections on vector bundles can be viewed as special cases of a  general Ehresmann connection. 

To specify an Ehresmann connection, one can specify a 1-form $\bomega$ on $E$ with values in the vertical sub-bundle $V={\rm ker}(d\pi)\subset TE$ such that for any section $v$ of $V$ one has $\bomega(v)=v$. Then the complementary horizontal sub-bundle $H$ of $TE$ is defined by the condition $\bomega(h)=0$, where $h$ is any section of $H\subset TE$. The horizontal sub-bundle has the property that $d\pi:TE\ra TS$ becomes a bundle isomorphism when restricted to $H$. Thus every vector field on the base $S$ can be lifted in a unique way to a horizontal vector field on $E$.

In our case, $S$ is the real line parameterized by the time coordinate $t$, while $F$ is a spatial slice with local coordinates $x^k$, so the 1-form $\bomega$ locally takes the form
\begin{equation}
       \boldsymbol{\omega}=\partial_{i}\otimes dx^{i}+N^{k}\partial_{k}\otimes dt.
\end{equation}
In other words, $\bomega$ is locally encoded in the connection coefficients $N^k({\bf x},t).$
The invariance of $\bomega$ under  a change of frame (\ref{framechange}) produces the transformation law (\ref{transformation}).

Given a function $h$ on space-time, one can define its covariant derivative $D_t h$ with respect to the time coordinate $t$ by lifting the vector field $\partial/\partial t$ to a horizontal vector field and taking the derivative of $h$ along this vector field. In coordinates, this gives 
\begin{equation}
D_t h=\partial_t h-N^j\partial_j h .
\end{equation}
The covariant derivative looks the same in all reference frames. 

Importantly, the connection coefficients $N^j$ can always be made to vanish locally by a suitable choice of frame. One seeks a reference frame change $x^k\mapsto \tx^k=\tx^k({\bf x}, t)$ such that the left hand side of equation (\ref{transformation}) is zero. To find the required frame change, the method of characteristics is employed and yields a system of non-linear ordinary differential equations
\begin{equation}
\label{connection}
\frac{dx^{j}(t)}{dt}=-N^{j}(\textbf{x}(t), t),\quad x^j(0)=\tx^j.
\end{equation}
The solution $x^j({\bf{\tx}},t)$ of the system (\ref{connection}) gives the the necessary transformation $\tx^{k}({\bf x}, t)$ implicitly.
A solution always exists locally in $t$. Globally a solution may fail to exist because a local solution $x^j({\bf{\tx}},t)$ may escape to infinity in a finite time. One may call a frame where $N^j$ is identically zero an inertial frame. If $N^j=0$ only in some region of space-time, then one is dealing with a locally inertial frame.

It is instructive to compare the above approach to the Newton-Cartan framework \cite{Cartan1,Cartan2,Kuchar,DuvalKunzle} (see also \cite{Obersetal,Bekaert&Morand,Geracie} for recent discussions). It describes the Newtonian gravitational field in terms of a spatial metric $\gamma^{\mu\nu}$ satisfying $n_{\mu}\gamma^{\mu\nu}=0$, where $n_{\mu}$ is a nowhere vanishing 1-form, a vector field $u^{\mu}$ with the constraint $u^{\mu}n_{\mu}=1$, and a 1-form $A_{\mu}$ encoding (part of) the Newtonian potential. Such a choice of $u^{\mu}$ is not unique; accordingly one requires invariance under \textit{Milne Boosts}:
\begin{equation}
    u^{\mu}\mapsto u^{\mu}+k^{\mu}, \hspace{1cm} A_{\mu} \mapsto A_{\mu}+\gamma_{\mu\nu}k^{\nu}-\frac{1}{2}\gamma_{\rho\sigma}k^{\rho}k^{\sigma}n_{\mu}, \hspace{1cm} k^{\mu}n_{\mu}=0,
\end{equation}
where $\gamma_{\mu\nu}$ is defined by the constraints $\gamma_{\mu\nu}u^{\nu}=0$ and $\gamma_{\mu\nu}\gamma^{\nu\rho}=\delta^{\rho}_{\mu}-n_{\mu}u^{\rho}$.
The Newton-Cartan formalism is related to ours through the dictionary:
\begin{equation}
\begin{split}
N^{k}&=A^{k}-u^{k},\\
\phi&=A_{\mu}(2u^{\mu}-A^{\mu}).
\end{split}
\end{equation}
It can be checked that $N^{k}$ and $\phi$ are Milne-invariant. In effect, the Newton-Cartan formalism contains redundant fields as well as an additional gauge symmetry (Milne boosts). Introducing the Ehresmann connection and $\phi$ eliminates this redundancy as well as the need for considering Milne boosts. This is explicitly stated in \cite{Bekaert&Morand}, where the connection coefficients $N^k$ were called a ``Coriolis-free field of observers", but is implicit in other works such as \cite{Kuchar,Obersetal}. 

\section{Classical and quantum particles in a gravitational field}

A covariant action for a particle of mass $m$ in a Newtonian potential $\phi$ takes the form
\begin{equation}
S=\frac{m}{2}\int dt\left[ \left(\frac{dx^j}{dt}+N^j\right)\left(\frac{dx^k}{dt}+N^k\right)h_{jk}    -2\phi \right].
\end{equation}
Here the spatial metric $h_{jk}=h_{kj}$ is positive-definite but otherwise may be an arbitrary matrix function of ${\bf x}$ and $t$.   Note that the Newtonian potential $\phi$ is invariant under the frame change (\ref{framechange}) and when $\phi=0$ and $N^j=0$ (that is, in the absence of the Newtonian potential and in an inertial frame) the action takes the standard form
\begin{equation}
S=\frac{m}{2}\int dt \left[\frac{dx^j}{dt}\frac{dx^k}{dt}h_{jk}\right].
\end{equation}
Thus a particle which was at rest at $t=0$ remains at rest at $t>0$. This is what distinguishes inertial frames from all other frames.

The key observation is that the above action is invariant (up to boundary terms) under the following transformation of the connection $N^j$ and the Newtonian potential $\phi$:
\begin{equation} \label{EPsymmetry1}
\begin{split}
N^{j} &\rightarrow N^{j} + h^{jk}\partial_{k}F,\\
\phi &\rightarrow \phi-D_{t}F+\frac12 h^{jk}\partial_{j}F\partial_{k}F.
\end{split}
\end{equation}
Such transformations depend on an arbitrary function $F$. This opens  possibility to eliminate the Newtonian potential at the expense of modifying the connection $N^j$. To do this, one has to solve the equation
\begin{equation}
\phi-D_{t}F+\frac12 h^{jk}\partial_{j}F\partial_{k}F=0.
\end{equation}
This equation reduces to the usual Hamilton-Jacobi equation for a particle in a Newtonian potential if we use an inertial frame where $N^j=0$. 

A solution to the Hamilton-Jacobi equation exists locally in $t$. Suppose we started from a nonzero Newtonian potential $\phi$ and zero $N^j$. Having solved the Hamilton-Jacobi equation for $F$, we can perform the transformations (\ref{EPsymmetry1}) and make $\phi=0$ at the expense of making $N^j$ non-zero. Then we can eliminate $N^j$ (again locally in $t$) by solving the equations (\ref{connection}) and performing the corresponding change of coordinates. This changes the metric too. The net result is that we eliminated $\phi$ at the expense of changing the spatial metric. 

The origin of the symmetry (\ref{EPsymmetry1}) can be traced back to time-reparameterization invariance which is present in the relativistic theory but not in the non-relativistic one.  To see this, note that once a global time coordinate has been chosen, a pseudo-Riemannian metric on space-time defines an Ehresmann connection via an ADM parameterization of the metric \cite{MTW}:
\begin{equation}\label{ADM}
ds^2=-c^2 \left(1+\frac{2\phi}{c^2}\right) dt^2+h_{jk} (dx^j+N^j dt)(dx^k+N^k dt).
\end{equation}
Here $c$ is the speed of light. The identification of the Newtonian potential $\phi$ in terms of the ADM ``lapse" function is standard \cite{MTW}. It is easy to check that under transformations of spatial coordinates (\ref{framechange}) the ``shift" vector field $N^j$ transforms as in (\ref{transformation}) and thus can be regarded as an Ehresmann connection. 
We are interested in the transformation of the function $\phi$ and the vector field $N^j$ under time-reparameterization $t=t'-\frac{F({\bf x},t')}{c^2}$. While it is complicated in general, it simplifies in the limit $c\ra\infty$. Keeping only the terms with non-negative powers of $c$, we find precisely the transformation (\ref{EPsymmetry1}). 

In a sense, the symmetry (\ref{EPsymmetry1}) is how the non-relativistic theory ``knows"  it arose as a limit of a relativistic one. This makes precise A. Einstein's guess that the Strong Equivalence Principle is explained by the   diffeomorphism invariance of the relativistic theory.

In the quantum case, it is easiest to start with a covariantized action for a Schr\"odinger field $\Psi$:
\begin{equation}\label{Ssch}
    S_{Schr}= \int  d^{3}x\,dt\, \sqrt h\,\left[\frac{i}{2}\left(\bar{\Psi}D_t\Psi-\Psi D_t\bar\Psi\right) - \phi m\bar{\Psi}\Psi-\frac{1}{2m}h^{jk}\partial_{j}\bar{\Psi}\partial_{k}\Psi\right],
\end{equation}
where $h={\rm det}\, ||h_{jk}||$. 
The Schr\"odinger equation derived from this action is
\begin{equation}\label{sch}
i D_t\Psi= -\frac{1}{2m\sqrt h}\partial_j \left(\sqrt h h^{jk} \partial_k\Psi\right)-m\phi\Psi+\frac{i}{2}\left[\partial_{k}N^{k}-D_{t}(\log{\sqrt{h}})\right]\Psi .
\end{equation}
This agrees with the results of \cite{Kuchar,DuvalKunzle} obtained using Newton-Cartan geometry. It can be checked that the equation (\ref{sch}) is covariant under frame-changes  (\ref{framechange}) if we postulate that the wave-function $\Psi$ transforms as a scalar. The total  probability $\int |\Psi|^2 \sqrt h d^3 x$ is also invariant under these transformations and is independent of time, ensuring a consistent physical interpretation of $|\Psi|^2$ as the probability density. It is also easily checked that the action (\ref{Ssch}) and the equation (\ref{sch}) are  invariant under the transformations (\ref{EPsymmetry1}) if we also transform the wave-function by a phase:
\begin{equation}\label{EPsymmetry2}
\Psi({\bf x},t)\mapsto e^{imF({\bf x},t)}\Psi({\bf x},t) .
\end{equation}
Thus in the quantum case it is also true that one can eliminate the Newtonian potential by changing a frame. 

Let us summarize the above discussion in  physical terms. In non-relativistic physics the effects of gravity on matter are described by a Newtonian potential together with a curved and time-dependent spatial geometry. The equivalence principle states that by changing a reference frame the Newtonian potential can always be eliminated locally at the expense of changing the spatial geometry.

\section{Examples}

In this section we consider a few examples of elimination of a Newtonian potential. We always start with a flat spatial metric, $d\ell^{2}=dx^2+dy^{2}+dz^{2}$.  

\subsection{Uniform gravitational field}
Let the Newtonian potential be $\phi=gz$. 
The Hamilton-Jacobi equation is
\begin{equation}
     \partial_{t}F=\frac{1}{2}(\nabla F)^2+gz.
\end{equation}
%The solution to this PDE takes the form:
%\begin{equation}
%    F(t,x,y,z)=Et+Ax+By+F_{z}(z),
%\end{equation}
%where $E$, $A$, and $B$ are arbitrary constants, the constant $E$ being identified with the energy. We set $A=B=0$ but keep the energy $E$ nonzero. The solution F is then
A particular solution of this PDE is
\begin{equation}
F=gzt+\frac{1}{6}g^2 t^3.
\end{equation}
The corresponding connection coefficients $N^j$ are $(0,0,gt)$. Solving the equation (\ref{connection}), we find that the following frame change  eliminates a uniform gravitational field:
\begin{equation}
\tilde z=z+\frac{gt^2}{2}.
\end{equation}
Since this transformation is a $t$-dependent isometry, the spatial metric is unchanged. This example is discussed in \cite{Kuchar}.

\subsection{Kepler problem}
In radial coordinates $(r,\theta,\varphi)$ the metric takes the form $d\ell^2=dr^2+r^2(d\theta^2+\sin^2\theta\, d\varphi^2)$. Consider the Newtonian potential  $\phi=-\frac{G}{r}$. The Hamilton-Jacobi equation is
\begin{equation}
 \partial_{t}F=\frac{1}{2}(\nabla F)^2-\frac{G}{r}.
\end{equation}
A particular solution of this PDE is
\begin{equation}
    F=\sqrt{8Gr}.
\end{equation}
The corresponding connection coefficients $N^{j}$ are $(\sqrt{\frac{2G}{r}},0,0)$.
A frame change which eliminates the connection $N^{k}$ is 

\begin{equation}\label{kepler_frame_change}
     \tr(r,t)=\left(r^{3/2}+t \sqrt{\frac{9G}{2}}\right)^{2/3}.
\end{equation}
In the new frame the metric becomes time-dependent:
\begin{equation}
d\ell^2=\frac{d\tr^2}{\left(1-\frac{t}{\tr^{3/2}} \sqrt{\frac{9G}{2}}\right)^{2/3}} + \left(1-\frac{t}{\tr^{3/2}} \sqrt{ \frac{9G}{2}}\right)^{4/3} \tr^2 (d\theta^2+\sin^2\theta\, d\varphi^2).
\end{equation}
One can show that this is unavoidable: any solution of the Hamilton-Jacobi equation which can be used to eliminate the Newton-Kepler potential $\phi=-G/r$ leads to a time-dependent spatial metric. Note also that the frame-change (\ref{kepler_frame_change}) is only defined locally. 

\subsection{Generalized Einstein elevators}

It is easy to determine which Newtonian potentials can be eliminated without affecting the flat spatial metric. The most general such potential has the form
\begin{equation}\label{elevator}
\phi({\bf x},t)={\bf x}\cdot \frac{d^2 {\bf w}(t)}{dt^2}-\frac12 \left(\frac{d{\bf w}(t)}{dt}\right)^2,
\end{equation}
where ${\bf w}(t)$ is an arbitrary vector-valued function of $t$. This formula describes a uniform but not static gravitational field. The corresponding solution of the Hamilton-Jacobi equation is $F={\bf x} \cdot \frac{d{\bf w}}{dt}$, and the frame-change is $\tilde {\bf x}={\bf x}+{\bf w}(t).$ This example is also discussed in \cite{Kuchar}.

%The spatial metric becomes
%\begin{equation}
%    ds^{2}=\frac{1}{r}(r^{\frac{3}{2}}-\sqrt{\frac{9}{2}G}t)^{\frac{2}{3}}dr^{2}+r^{2}d\Omega^{2}.
%\end{equation}    
%The solution to this PDE takes the form:
%\begin{equation}
%F(t,r,\theta,\phi)=Et+p_{\phi}\phi + F_{r}(r)+F_{\theta}(\theta),
%\end{equation}
%where $p_{\phi}$ and $E$ (the energy) are constants. The separation of variables %yields two ODE's:
%\begin{equation}
%    (\partial_{\theta}F_{\theta})^{2}+\frac{p_{\phi}^2}{\sin^{2}\theta}=\beta,
%\end{equation}
%\begin{equation}
%    \frac{1}{2}(\partial_{r}F_{r})^2-\frac{C}{r}+\frac{\beta}{2r^2}=E,
%\end{equation}
%where $\beta$ is an arbitrary constant. For simplicity, we set $E=\beta=p_{\phi}=0$, %&or more of $E$, $\beta$, and $p_{\phi}$ are nonzero. The solution F is then,
%\begin{equation}
%F(t,r,\theta,\phi)=\int dr\sqrt{2\frac{C}{r}}=\sqrt{8C}(\sqrt{r}-\sqrt{r_{0}}).
%\end{equation}
%The coordinate change which eliminates the connection $N^{k}$ is
%\begin{equation}
%     \tilde{r}(r,t)=(r^{\frac{3}{2}}-\frac{3}{2}\sqrt{2C}t)^{\frac{2}{3}}.
%\end{equation}
%The spatial metric becomes \begin{equation}h_{ij}=
%      \begin{pmatrix}
%     0 & 0 & 0 & 0 \\
%    0 & \frac{1}{r}(r^{\frac{3}{2}}-\frac{3}{2}\sqrt{2C}t)^{\frac{2}{3}} & 0     & 0 %\\
%     0 & 0     & r^2 & 0 \\
%     0 & 0     & 0     & r^2\sin^{2}(\theta)           
%    \end{pmatrix}.
%\end{equation}

\section{Discussion}

A student first studying General Relativity is often confused by the statement that physics should be formulated in a frame-independent way. Does it mean that one has to go back and re-examine Newtonian physics and non-relativistic quantum mechanics to ensure that they conform to this principle? The above discussion shows that the required modification of non-relativistic physics is fairly  straightforward and mostly amounts to replacing time derivatives with covariant time derivatives (with respect to the Ehresmann connection) and allowing for a general spatial metric. The usual equations are recovered by going to an inertial frame where the  connection coefficients $N^j$ vanish. However, the covariant approach shows that thanks to the symmetry (\ref{EPsymmetry1}) there is an ambiguity in the definition of $N^j$ and therefore an ambiguity in the definition of the class of inertial frames. It is this ambiguity that underlies the equivalence principle. For several species of particles, the symmetry (\ref{EPsymmetry1}) is present only if the gravitational mass is equal to the inertial mass. This provides a symmetry-based explanation of the equivalence principle in a  non-relativistic setting. 

The geometric approach also clarifies the transformation properties of the wave-function under frame changes. As explained in many textbooks (see e.g. \cite{LL}), the wave-function transforms under Galilean boosts with a non-obvious phase factor:
\begin{equation}
\Psi({\bf r},t)\mapsto \Psi({\bf r}-{\bf v} t,t) e^{i m\left({\bf v}\cdot{\bf r}-\frac12 {\bf v}^2 t\right)}.
\end{equation}
It is also well-known that transformation to a uniformly accelerated frame involves a phase factor of the form $\exp(i m (t {\bf a}\cdot {\bf r}-a^2 t^3/6)).$ This seems to suggest that such phase factors will also be required when transforming to a uniformly rotating or a general non-inertial frame. As as explained above, the wave-function transforms as a scalar under arbitrary changes of the reference frame. Non-trivial phase factors appear when one performs  non-geometric transformations (\ref{EPsymmetry1},\ref{EPsymmetry2}). In certain cases, such as Galilean boosts and uniformly accelerated frames, these transformations can be used to eliminate the connection coefficients $N^j$ generated by the frame change. It is these transformations, not frame changes, which generate  non-obvious phase factors for the wave-function.

Leaving aside gravity, the version of the non-relativistic geometry discussed in this paper should be useful in other areas of physics. Recently there has been a renewed interest in the Effective Field Theory approach to hydrodynamics, see \cite{review} for a review. In this approach, hydrodynamics in a broad sense is described by a sigma-model whose form is constrained by symmetries of the particular problem. The geometric framework discussed in this paper should be a suitable starting point for very general non-relativistic theories of hydrodynamic behavior, including the cases without either Galilean or spatial translation  symmetry.

\noindent
{\bf Acknowledgements:}
M. T. is grateful to the Caltech SURF program for providing a research opportunity. This research was supported in part by the U.S.\ Department of Energy, Office of Science, Office of High Energy Physics, under Award Number DE-SC0011632. A.K. was also supported by the Simons Investigator Award. \\

\end{document}